# Ultrafast high-temperature sintering of dense and textured alumina

Rohit Pratyush Behera[1], Matthew Jun-Hui Reavley[2,3], Zehui Du[3], Gan Chee Lip[2,3], Hortense Le Ferrand[1, 2*]

[1] *School of Mechanical and Aerospace Engineering, Nanyang Technological University, 50 Nanyang avenue, Singapore 639798*

[2] *School of Materials Science and Engineering, Nanyang Technological University, 50 Nanyang avenue, Singapore 639798*

[3] *Temasek Laboratories, Nanyang Technological University, 50 Nanyang Drive, Singapore 637553*

* Corresponding authors: hortense@ntu.edu.sg

## Abstract

Crystallographic texture engineering in ceramics is essential to achieve direction-specific properties. Current texture engineering methods are time-consuming, energy extensive, or can lead to unnecessary diffusion of added dopants. Herein, we explore ultrafast high-temperature sintering (UHS) to prepare dense and textured alumina using templated grain growth (TGG). From a slurry containing alumina microplatelets coated with $Fe_3O_4$ nanoparticles dispersed in a matrix of alumina nanoparticles, green bodies with oriented microplatelets were prepared using magnetic assisted slip casting (MASC). The effects of the sintering temperature, time and heating rate on the density and microstructure of the obtained ceramics were then studied. We found that TGG occurs for a temperature range between 1640 and 1780 °C and 10 s sintering time. Sintering at 1700 °C for 10 s led to dense and textured alumina with anisotropic grains thanks to the $Fe_3O_4$ coating, which did not have the time to diffuse. The highest texture and relative density were obtained with a heating rate of ~5,500 °C/min, leading to texture-dependent anisotropic mechanical properties. This study opens new avenues for fabricating textured ceramics in ultra-short times.

## 1. Introduction

Crystallographic texture engineering in ceramics is essential to achieve direction-specific properties, such as mechanical, thermoelectric, electronic, optical, piezoelectric, and magnetic [1–3]. Crystallographic texture control in bulk ceramics is typically achieved using ultra-high magnetic fields to orient isotropic grains along a preferential direction [4], grain morphology engineering where anisotropic single-crystal grains are aligned by a variety of methods, such as external field-assisted techniques [5], uniaxial pressing [6], tape casting [7], freeze casting [8], or templated grain growth (TGG). In TGG, a small concentration of anisotropic grains is distributed in a matrix of nanoparticles to serve as crystallographic seeds for the nanoparticles during sintering. TGG can be regulated by tuning the amount and size of initial anisotropic templates and nanoparticles and adding dopants such as flux, liquid precursors, and other constituents [9–17]. Contrary to the other processes, TGG has the advantage of being pressureless. However, TGG requires a long sintering time, which is time-consuming and energy-intensive.

A solution to accelerate the TGG could be to increase the heating rate during sintering [25]. Additionally, fast heating rates could offer benefits such as limiting the grain growth and retaining fine grain microstructure [18], promoting densification [19], and reducing the time for reaction and volatilisation in solid-state batteries consisting of volatile elements like Li, Na [19]. Current fast heating rate sintering methods are spark plasma sintering (SPS), flash sintering (FS) and microwave sintering (MW) etc. [18,20]. Although SPS has been used in conjunction with TGG [21,22], the sintering of dense and textured alumina *via* TGG with a very-fast heating rate has not been reported to date.

Recently, a pressureless sintering method called ultrafast high-temperature sintering (UHS) has been developed, which utilises conduction and radiation heating to sinter the ceramics within seconds [19]. Many works done using UHS have achieved good densification of ceramics with various compositions [23–27]. Although the ultra-high heating rates up to 20,000 °C/min in UHS limit diffusion, grain growth is still observed [19,28,29]. To the best of the author's knowledge, the use of UHS to produce ceramics with crystallographic texture control remains unexplored.

In this work, we aim to fabricate dense and textured alumina ceramics within seconds using UHS and TGG. The approach consists in producing green bodies with magnetically-aligned alumina templates. To do so, alumina microplatelets were coated with $Fe_3O_4$ nanoparticles and dispersed in a matrix of alumina nanoparticles. This method has been proven useful in controlling the local orientation of grains in ceramics for fine-tuning the properties [30]. We study the influence of the sintering temperature, sintering time, and heating rate on the density and grain size of the ceramics obtained after UHS. This allows us to draw a sintering phase map and to identify the conditions at which TGG occurs, leading to high relative density and texture. $Fe_3O_4$ coating was also looked at to understand its role in maintaining the anisotropy of the grains after sintering. Finally, we tested the mechanical properties at nano and micro scales to observe the effect of crystallographic texture in dictating anisotropic properties. The results from this work could be applied to other ceramic systems for texture development and property enhancement using UHS.

## 2. Experimental procedure

### 2.1 Fabrication of textured alumina

*2.1.1 Magnetically responsive microplatelets*

To make alumina microplatelets (10030, diameter ~6.36 μm, thickness ~0.25 μm, aspect ratio ~ 25-30, Kinsei Matec) magnetically responsive, we adsorbed superparamagnetic iron oxide nanoparticles (SPIONs) (10 nm, EMG-705, Ferrotec) on their surface following an established procedure [31]. At first, the microplatelets were dispersed in a large volume of deionised water, followed by the addition of 0.75 vol% SPIONs with respect to the microplatelets. Then, the mixture was left stirring until complete electrostatic adsorption of the SPIONs nanoparticles on the surface of the alumina microplatelets. We obtained the final functionalised microplatelets by filtration followed by drying overnight in a heating oven (IKA Oven 125) at 40 °C. The successful adsorption of the SPIONs on the surface of the microplatelets was verified by the brownish appearance of the magnetised powders. Also, the SPIONs were visible by electron microscopy (see inset in **Fig 1a**). The surface coverage of $Fe_3O_4$ nanoparticles on the microplatelets was measured from the electron micrographs to be around ~10-15% using Image J.

*2.1.2 Preparation of the slurry*

The slurry consisted of a bimodal colloidal system with the magnetised alumina microplatelets dispersed in water with alumina nanoparticles. The microplatelets to nanoparticles ratio was 5 to 95 and the total solid loading was 55%. The alumina nanoparticles used in the mixture ranged from 100-120 nm as a commercial suspension (AERODISP® W 440, Evonik, ~120 nm) was supplemented with solid nanoparticles (AP-D powder, Struers, ~100 nm) to attain the desired solid loading. All the constituents were mixed using a vortex mixer, then using probe ultrasonication (Sonopuls Serie 4000, Bandelin, pulse-on = 1 s, pulse-off = 2 s) at 13% amplitude. The suspensions were cooled each time after the mixing. The process was repeated three times to ensure homogeneous mixing and dispersion.

*2.1.3 Magnetically assisted slip casting of the green bodies*

The bimodal colloidal slurry was casted in plastic moulds of 1 cm height that were glued onto a gypsum substrate (Ceramix). The mould was positioned under a neodymium magnet (RS component) rotating at ~ 1 Hz. The field strength at the place of the sample was 50-75 mT. The rotation of the magnet was conducted in order to obtain horizontal alignment of the basal or c-plane of the magnetised microplatelets (**Fig. 1a**). The magnetic field strength and its rotation were maintained until all the water from the slurry got removed through the porous gypsum substrate. Before unmoulding, the casted samples were dried overnight at 40 °C. The obtained green bodies had a relative density of ~ 52 $\pm$ 2.07 %.

*2.1.4 Sintering of green bodies*

The green bodies were sintered using pressureless processes: conventional sintering (CS) and ultrafast high-temperature sintering (UHS), as described below, to compare the two different sintering processes with different heat transfer mechanisms.

*2.1.4.1 Conventional sintering (CS)*

The CS of the obtained green bodies was done in a high-temperature furnace (Nabertherm, High-temperature furnace LHTCT01/16). The samples were first heated at 500 °C for 5 h. Then, the sintering was done in air at a heating rate of 2.5 °C/min, with a dwell time of 2 h and 10 h at a sintering temperature of 1600 °C, followed by natural cooling.

*2.1.4.2 Ultrafast high-temperature sintering (UHS)*

UHS uses conduction and radiation heating to sinter ceramics [19]. A schematics of the setup is presented in **Fig. 1b**. To achieve the desired temperature-time profile for sintering, we regulated the power from the supply unit (Delta Elektronika Sm1500 70-22) to the heating section using a microcontroller by adjusting the current $i$ operated by a computer. The temperature-time profile was defined and programmed to regulate the desired power needed from the supply unit. The microcontroller regulated the power output, keeping the heating temperature constant (see **supplementary note 1** for explanation). The output power was then applied across the carbon fabric (AvCarb HCB) of fixed dimensions, 10 cm in length and 5 mm in breadth, connected *via* copper contacts to induce heat into the sample. The green bodies were kept within 1 mm thickness and 2-3 mm in the lateral dimension to prevent the formation of a temperature gradient throughout their thickness. The entire heating setup was placed in a glove box with a 99.9% pure argon atmosphere, and the temperature was measured using a pyrometer (Raytek Raynger 3i +, measuring range: 700-3000 °C, precision $\pm$ 10°C). The pyrometer was aimed at the top of the carbon fabric, directly in contact with the sample. The temperature difference between the top and the bottom surface of the ceramic is expected to be at less than 15 °C after 5 s, thanks to the low thickness of the green bodies (less than 1 mm based). Indeed, previous work showed that there was less than 20 °C difference for a 2 mm thick barium titanate sample placed inside an alumina crucible within 6 s of the ramp-up stage, [32]. This difference was found to be less than 5 °C after 10 s. Here, we regulated the input current $i$ to control the following sintering parameters, namely the sintering temperature $T_s$, the sintering time $t_s$, and the heating rate $r_s$ (**Fig. 1c**). The values used for the sintering parameters are listed in **Table 1**. A range of heating rates are given instead of single values to accommodate the change in $T_s$ within that particular range.

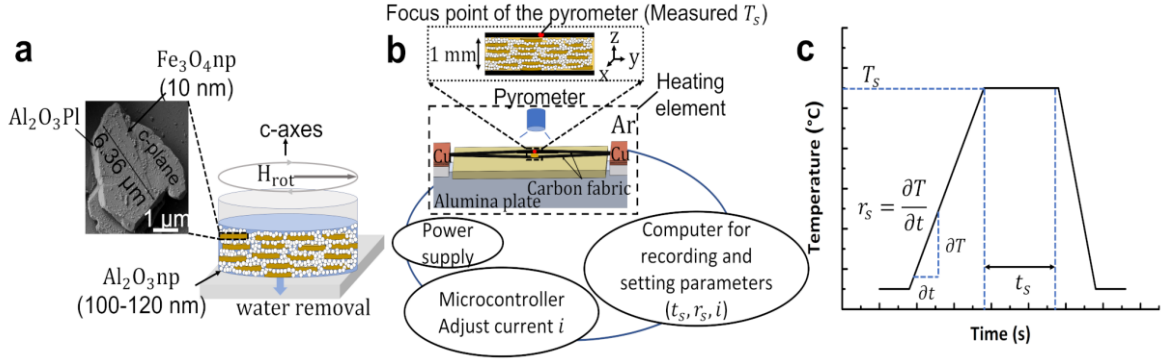

**Figure 1: Preparation of the oriented green bodies using MASC and UHS process. (a)** Schematics of the MASC process applied to an aqueous slurry composed of magnetically responsive alumina microplatelets (pl) decorated with $Fe_3O_4$ nanoparticles (np) and suspended with alumina nanoparticles (np). A magnetic field ($H_{rot}$) of 50-75 mT and rotating at a frequency of ~1 Hz aligns the c-plane of the microplatelets horizontally along the rotational axis of $H_{rot}$. The crystallographic c-axis is perpendicular to the c-plane of the microplatelets. **(b)** Flow chart of the UHS process. The power supply is regulated by adjusting the current $i$ from the microcontroller to obtain the desired temperature at the sample placed between the carbon fabric. A pyrometer measures the sintering temperature $T_s$ of the top surface of the fabric, directly in contact with the sample as illustrated with the red dot, while a computer records the exact isothermal sintering time $t_s$ of the sample. **(c)** Temperature-time plot showing the sintering parameters that were varied in this study: sintering temperature $T_s$, isothermal sintering time $t_s$, and heating rate $r_s$.

**Table 1:** Values of the sintering parameters varied in this study.

|  | $T_s$ (°C) | $t_s$ (s) | $r_s$ (°C/min) |
|---|---|---|---|
| experiment 1 | 1600, 1650, 1700, 1800 | 10 (fixed) | 10000-12000 (fixed) |
| experiment 2 | 1700 (fixed) | 10, 15, 20 | 10000-12000 (fixed) |
| experiment 3 | 1700 (fixed) | 10 (fixed) | 2000-3000, 5000-6000, 10000-12000 |

## 2.2 Materials Characterisation

*2.2.1 Density characterisation*

The Archimedes principle was used to measure the density $\rho_s$ of the ceramics after overnight impregnation with ethanol, using:

$$\rho_s = \rho_{Eth} \times \frac{W_A}{W_{IA}-W_{IW}},$$

where $\rho_{Eth}$ is the density of ethanol at room temperature (0.789 g/cm³), $W_A$ is the dried weight of the ceramic, $W_{IA}$ is the weight of the impregnated ceramic measured in air and $W_{IW}$ is the weight of the impregnated ceramic measured in ethanol.

The relative density (RD) in % was calculated using:

$$\frac{\rho_s}{\rho_{fully\ dense}} \times 100,$$

with $\rho_{fully\ dense}$ = 3.95 g/cm³. At least 5 samples were measured.

*2.2.2 Microstructure, phase detection and elemental characterisation*

We observed the fractured surface of the samples using a field emission scanning electron microscope (FESEM, JOEL 7600F) after depositing a thin layer (~10 nm) of gold on the surface. The crystallographic phases present in the samples were identified using X-ray diffraction (Bruker D8 Advance X-ray diffractometer) with Cu Kα radiation in the Bragg-Brentano geometry. The spectra were recorded for a range of 2θ varying from 10 to 60 ° with a step size of 0.03 °. Elemental mapping (EDX) was used to characterise the distribution of the elements in the magnetised microplatelets, green body, CS samples and UHS samples using an electron microscope (JEOL, 7800 F) at 20 kV after gold coating. The EDX mapping for the UHS samples was performed without coating.

*2.2.3 Crystallographic orientation, band contrast images and estimation of grain dimensions*

The ceramics were mirror-polished using SiC papers (320P, 500P, 800P, 1200P and 2400P, Struers), followed by colloidal suspensions (AP-D solution, 0.3 μm diameter Struers and OP-U solution, 0.04 μm diameter, Struers). Crystallographic mappings were carried out using Electron Backscatter Diffraction (EBSD) on uncoated samples after polishing, using an electron microscope (JEOL, 7800 F) at 20 kV and a step size of 0.21 μm. Band contrast images were acquired using Aztec crystal software. The bright and dark regions represent good and poor EBSD patterns, respectively. The grain diameter *d* and aspect ratio *s* were estimated from the EBSD patterns taken on the x-y and z-y planes of the samples (see inset of **Fig. 1b** for axes definition) using the Aztec crystal software. The grain sizing settings were 10 pixels per grain, grain detection angle of 10 ° and threshold angle of 10 °.

## 2.3 Mechanical characterisation

*2.3.1 Young's modulus*

Young's moduli were measured using a Berkovich tip mounted nanoindenter (G200, KLA Tencor) on the mirror polished samples. Twenty indents were made with a spacing of 50 µm between each indent. The tests were performed below a thermal drift of 0.1 nm/s under load-controlled conditions. The maximum load was 200 mN, and the loading rate was kept at 1 mN/s for an indentation depth of at least 600 nm to estimate the moduli values accurately. The unloading was performed till 90% of the loading to avoid any effects of the thermal drift on the unloading curve. The peak hold time and the Poisson's ratio were set at 10 s and 0.22, respectively. Oliver-Pharr's theory was used to obtain the Young's modulus values from the unloading curve of each indentation. The reliability of the measured values was verified using MATLAB (R2019a) to ensure that Weibull's modulus was superior to 3.

*2.3.2 Vickers hardness*

The hardness was measured using a Vickers micro-indenter (Future-Tech, FM- 300E) on mirror polished samples at a load of 2 kgf and a dwell time of 10 s. The indents were then observed under FESEM (JOEL 7600F) after gold coating, and the diagonal lengths were measured using Image J to calculate the hardness for each indent. Ten indents were made under the same load to ensure reliability without the indentation size effect.

# 3. Results and Discussion

3.1 Effect of sintering parameters on density and grain morphology.

In this section, we study the effects of the sintering parameters $T_s$, $t_s$ and $r_s$, on the density, microstructure and grain dimensions of horizontally aligned alumina green bodies while keeping the cooling rate constant at 10000-12000 °C/min. In short, we conducted three consecutive experiments to determine the sintering parameters leading to dense and textured alumina *via* TGG using UHS. In the first experiment (experiment 1), the effects of $T_s$ are studied keeping $t_s$ and $r_s$ constant at 10 s and 10000-12000 °C/min, respectively. As per the outcome of experiment 1, the $T_s$ for the highest relative density and grain anisotropy is selected and used in the two other experiments. In the second experiment (experiment 2), $T_s$ and $r_s$ are fixed at the value obtained in experiment 1 and 10000-12000 °C/min respectively, while $t_s$ is varied. The $t_s$ that leads to highest relative density, and grain anisotropy is then used as input for the next experiment. So, in the third experiment (experiment 3), $T_s$ and $t_s$ are fixed at the values obtained in experiments 1 and 2, while $r_s$ is varied. The outcome of these experiments is thus a set of three sintering parameters that yield dense alumina with horizontally oriented anisotropic grains. Although we aim at high crystallographic texture, we hypothesise that the morphological grain microstructure and the crystallographic texture are related in our system, as pointed out in other works using alumina microplatelets [12,30]. Indeed, the alumina microplatelets employed here are single crystals with a well-defined basal or c-plane (**Fig. 1a,** inset). The crystallographic texture of the ceramics will be further confirmed with EBSD measurements in section 3.3 and Supplementary Information (SI) of this paper.

*3.1.1. Effect of sintering temperature (experiment 1)*

To observe the effects of $T_s$ only on the density and microstructure, we fixed $t_s$ at 10 s and $r_s$ at 10000-12000 °C/min and varied the $T_s$ with four different temperatures: 1600 °C, 1650 °C, 1700 °C and 1800 °C (see **supplementary Fig. 1** for the regulated power and temperature profiles). The $T_s$ chosen are close to alumina's solid-state sintering temperature, 1500-1800 °C [28], to better understand the densification and grain growth processes across these temperatures. All the obtained ceramics densified above 85 vol% despite the ultra-low sintering time of 10 s. The relative densities increased with the $T_s$, reaching a maximum of about 92 % at 1700 °C, above which no significant densification was observed (**Fig. 2a**). This behaviour of constant final relative density beyond a certain $T_s$ is also observed in SPS [33,34], and other sintering processes [35].

Contrary to the final relative densities, the microstructures of the UHS alumina obtained at these different temperatures varied significantly, showing differences in grain morphology (**Fig. 2b**). At $T_s$=1600 °C, the nanoparticles and the microplatelets could be observed distinctly. This is due to the ultra-short sintering time, which likely reduced the diffusion of atoms, limiting the grain growth while densifying. Nevertheless, some interparticle necking occurred, probably due to the nanoparticles' high surface area and curvature (**Fig. 2b**, 1600 °C) [36]. At $T_s$=1650 °C and 1700 °C, bigger anisotropic grains could be observed, suggesting the templated growth of the nanoparticles onto the microplatelets (**Fig. 2b**). This could be due to the melting of the $Fe_3O_4$ nanoparticles covering the microplatelets, whose melting point is ~ 1600 °C. These $Fe_3O_4$ nanoparticles likely melted under these sintering conditions and promoted the templated grain growth. At $T_s$=1800 °C, the grains look inhomogeneous, with some grains having abnormal ultra-large sizes. This is probably due to the partial melting of the alumina nanoparticles and microplatelets, which led to abnormal grain growth (AGG). This is confirmed by the absence of additional phases in the XRD pattern apart from alumina and Fe-containing phases (see **supplementary Fig. 2**), indicating no significant reaction between our carbon fabric and the sample due to the ultra-short sintering time, which minimises the side reactions. It is also interesting to note that all microstructures at the respective temperatures showed the expected horizontal alignment of the grains except at 1800 °C, where vertical features are apparent.

The effects of $T_s$ on the microstructure can be visualised by measuring the grain dimensions at these conditions (**Fig. 2c**) (see **supplementary Fig. 3** for grain diameter distributions). The mean grain diameter is the arithmetic mean of all grain diameters in the ceramics, as obtained from EBSD on polished cross-sections. The grain sizes measured appeared smaller than what is shown in the electron micrographs as EBSD reveals the grain boundaries more accurately (see **supplementary Fig. 4**). As the sintering occurs, and the texture develops *via* TGG, the grain diameter and aspect ratio are expected to increase. Here, grain diameter increased with $T_s$ linearly up to 1700 °C, from 4.45 μm to 5.2 μm, but the increment is not very large (**Fig. 2c**). However, the aspect ratio of the samples nearly tripled. Indeed, at 1600 °C, the isotropic nanoparticles still present led to a small overall aspect ratio. At 1650 °C, the aspect ratio increased due to TGG and continued rising until 1700 °C. This is due to the radial growth of templated grains, whereby the increase in aspect ratio is dominated by an increment in grain diameter compared to templated grain thickness which remains relatively constant (see **supplementary Fig. 5** for grain thickness comparison) [9].

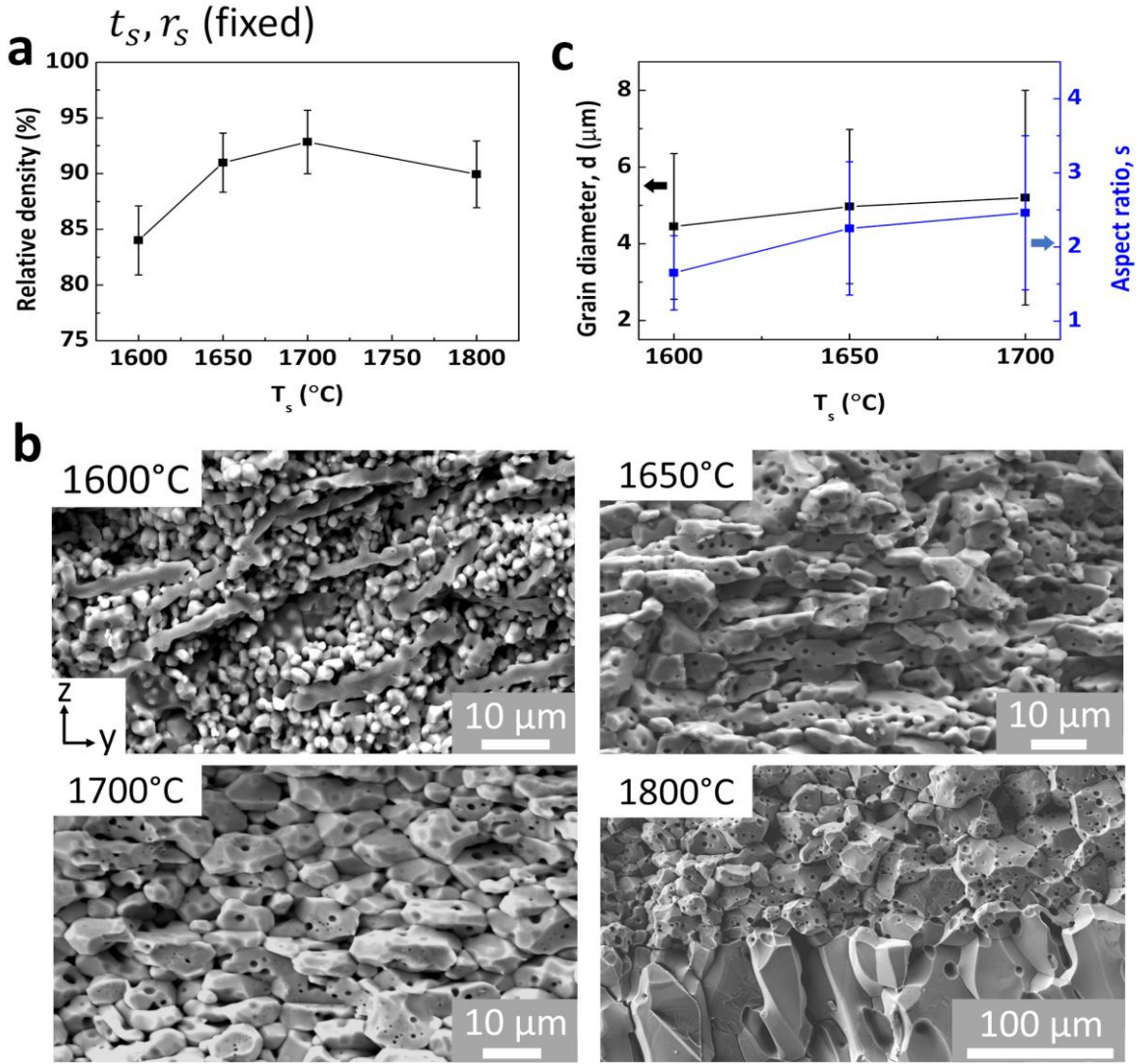

**Figure 2: Effect of variation of sintering temperature $T_s$ on density, microstructure and grain dimensions at a fixed sintering time $t_s$ of 10 s and heating rate $r_s$ of 10000-12000 °C/min. (a)** Relative density as a function of $T_s$. **(b)** Electron micrographs of the fractured cross-sections in the z-y plane (see inset in **Fig. 1b** for plane directions) for different $T_s$. **(c)** Grain diameter $d$ (in black) and aspect ratio $s$ (in blue) as a function of $T_s$. The inhomogeneous grain structure at 1800 °C prevented the determination of the grain size.

At fixed $t_s$ and $r_s$, the $T_s$ significantly affects the relative density, microstructure and grain dimensions. There is a domain of temperature from 1650 to 1700 °C, at which TGG occurs, leading to anisotropic grains and high relative density. At 1700 °C, we find high densification with a good aspect ratio. However, additional information is needed to understand the role of other sintering parameters $t_s$ and $r_s$ to achieve high densification with suitable anisotropy.

Therefore, in the next section, we study the effect of $t_s$ keeping the $T_s$ at 1700 °C and $r_s$ at 10000-12000 °C/min.

*3.1.2 Effect of sintering time (experiment 2)*

At fixed $T_s$=1700 °C and $r_s$=10000-12000 °C/min, we varied the $t_s$ for 10, 15 and 20 s and characterised the density, microstructure, and grain dimensions of the obtained ceramics. For lower values of $t_s$ such as 5 s, no TGG was observed, and the nanoparticles were still visible; therefore, we used a $t_s$ higher than 5 s. The relative density did not vary drastically with increasing $t_s$ (**Fig. 3a**). Surprisingly, some decrease in the density was observed above 10 s; however, it is not that significant. This decrease in density could be due to process-induced defects such as a large temperature hysteresis between the heating source and the sample due to the very high heating rate [37]. Additionally, at longer sintering times, beyond 10 s, it is hypothesised that the grains grew predominantly, trapping the pore between large grains and resulting in larger distances for transport of matter to close the pores [38,39]. At 20 s, the relative density was nearly equal to that of 15 s, which is ~90 %. This independence of the relative density with $t_s$ is also observed in other sintering processes with or without pressure [33–35,40,41].

The microstructures obtained exhibited the expected horizontal alignment, with a variation in the grain anisotropy (**Fig. 3b**). At 10 s, the grains appear to be more anisotropic than those obtained after 15 s and 20 s of sintering. Although this may not be striking from the electron micrographs, it is revealed in the grain morphology measurements, where the aspect ratio decreased from ~2.4 to ~1.5 (**Fig. 3c**). At 10 s, the interfacial $Fe_3O_4$ nanoparticles might have melted sufficiently to trigger TGG. However, at 15 and 20 s, the melting of $Fe_3O_4$ might have been large enough to facilitate the diffusion of atoms making the grains grow in thickness and decreasing their anisotropy. Indeed, it has been reported that sintering alumina in the presence of oxide of Fe generally inhibits grain growth along the thickness of alumina templates [42,43]. However, sintering in a reducing atmosphere could enhance diffusion due to the conversion of $Fe^{3+}$ to $Fe^{2+}$ [42,44]. Therefore, at $t_s$=10 s, it is likely that there was not sufficient time for the $Fe_3O_4$ to melt completely or diffuse and that they stayed at the grain boundary. Therefore, the alumina grains remained anisotropic and of small size (**Fig 3b,c**). However, at increased $t_s$ beyond 10 s, it is likely that the $Fe_3O_4$ had enough time to melt or diffuse, leaving the grain boundary free for the alumina grains to grow. Consequently, the grains obtained had a larger size but a lower anisotropy.

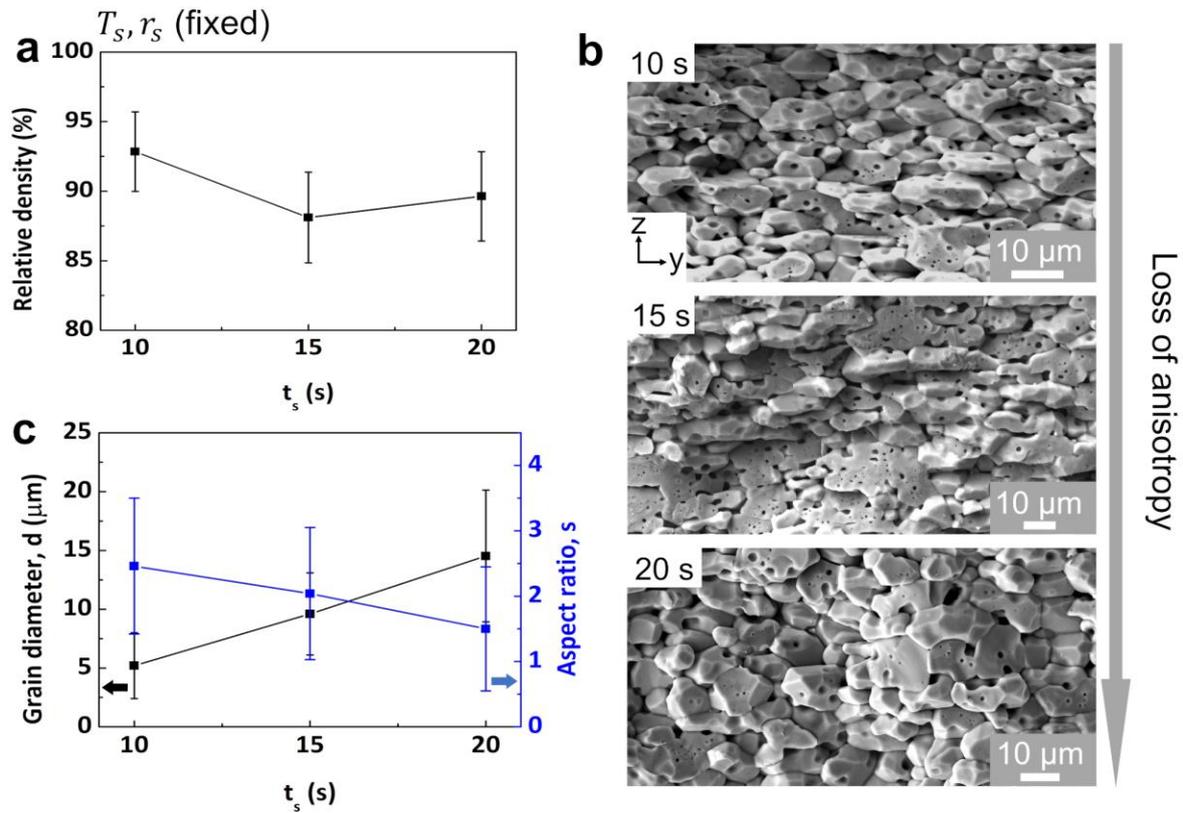

**Figure 3: Effect of the sintering time $t_s$ on the density, microstructure and grain dimensions at a heating rate $r_s$=10000-12000 °C/min and a sintering temperature $T_s$=1700 °C. (a)** Relative density as a function of $t_s$. **(b)** Electron micrographs of the fractured cross-section of the ceramics showing a decrease in the grain anisotropy with increasing $t_s$. **(c)** Grain diameter (in black) and aspect ratio (in blue) as a function of $t_s$.

To test the influence of the presence of $Fe_3O_4$ in the green body, we prepared green bodies made from a slurry composition containing alumina microplatelets coated 'with $Fe_3O_4$' and another containing alumina microplatelets 'without $Fe_3O_4$' (**Fig. 4**). Initially, we used the $Fe_3O_4$ nanoparticles to orient the alumina microplatelets horizontally in the green body. In the absence of $Fe_3O_4$, horizontal alignment was also observed due to gravitational forces, but $Fe_3O_4$ was found to be crucial in maintaining grain anisotropy after sintering. Here we used the sintering conditions of $T_s$=1700 °C, $t_s$=10 s and $r_s$=10000-12000 °C/min. For the ceramics made from the composition 'without $Fe_3O_4$', the grains were primarily equiaxial with a large diameter of ~21.2 μm. In contrast, the grains were anisotropic with a small diameter of ~5 μm for the ceramics made from the composition 'with $Fe_3O_4$'. The $Fe_3O_4$ coating, therefore, plays a pivotal role in our ceramics' microstructures. The grain diameter in the ceramics 'with $Fe_3O_4$' increased with increasing $t_s$ from ~5 μm at 10 s to ~15 μm at 20 s (**Fig. 3c**). However, this increase in the grain diameters coincides with a decrease in the aspect ratio.

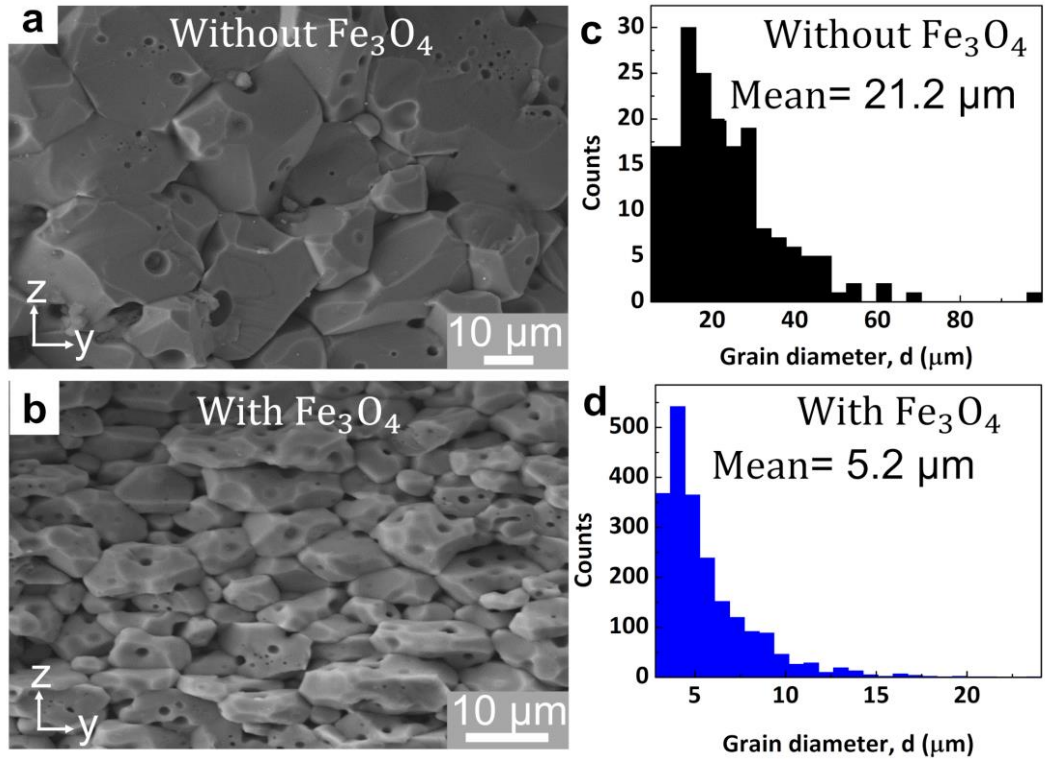

**Figure 4: Effect of $Fe_3O_4$ coating on the grain morphology after sintering at $r_s$=10000-12000 °C/min, $T_s$= 1700 °C for $t_s$= 10 s**. (**a**),(**b**) Electron micrographs of fractured cross-sections of samples made 'without $Fe_3O_4$' and 'with $Fe_3O_4$', showing equiaxial and anisotropic grains respectively. (**c**),(**d**) Grain size distribution for the ceramics 'without $Fe_3O_4$' and 'with $Fe_3O_4$' samples respectively. $d_m$ is the mean grain diameter.

The variation of $t_s$ at fixed $T_s$ and $r_s$ has a lesser effect on the relative density, but the grain anisotropy and diameter are significantly affected. At $t_s$ of 10 s, we found good density with the highest anisotropy despite lower grain diameter. Based on the results obtained so far, we therefore now fix the $t_s$ at 10 s and $T_s$ at 1700 °C to look at the effects of $r_s$ on the densification and the microstructure of the ceramics.

*3.1.3 Effect of heating rate (experiment 3)*

We varied the $r_s$ for three different ranges of heating rates, 2000-3000, 5000-6000 and 10000-12000 °C/min, keeping $T_s$ at 1700 °C and $t_s$ at 10 s, and characterised the relative density and microstructures. The relative density was significantly affected by the $r_s$ (**Fig. 5a**). By increasing $r_s$ from 2000-3000 to 5000-6000 °C/min, the relative density increased to ~98 %. This increase in the relative density with increasing in $r_s$ is expected due to increased mass diffusion [38]. However, when $r_s$ was increased from 5000-6000 °C/min to 10000-12000 °C/min, the densification reduced. One reason could be that at this extremely high $r_s$ of 10000-12000 °C/min, the ramp-up time where the temperature increases shortens, therefore reducing the overall sintering time. At these conditions, the $T_s$=1700 °C may not be sufficient enough for high densification [45,46]. Moreover, the relative density at $r_s$=10000-12000 °C/min was slightly lower than that obtained at 2000-3000 °C/min, likely due to insufficient time and energy for particle rearrangement and probable large temperature hysteresis between the heating element and the sample [37].

The morphology of the anisotropic grains across the different $r_s$ is shown in the band contrast images (**Fig. 5b**). The grains obtained at $r_s$ of 2000-3000 °C/min had an aspect ratio of ~2.6, which increased to ~3 when $r_s$ reached 5000-6000 °C/min (**Fig. 5c**). This is because comparably larger grains developed due to enhanced diffusion, with grain diameter increasing from 6.5 to 8 µm (**Fig. 5c**) (see **supplementary Fig. 6** grain diameter distributions). However, the aspect ratio decreased when $r_s$ was further increased to 10000-12000 °C due to the smaller grain diameter to ~ 5.2 µm. We found the aspect ratio to be dependent mainly on the grain diameter as the grain thickness stays relatively constant for the tested $r_s$. Indeed, the grain thickness was found to vary between 2.10-2.62 µm only (see **supplementary Fig. 7** for grain thickness distribution). This is in good agreement with the fact that in TGG, the radial growth of templates occurs primarily [9,47]. However, it is also noticeable that the grain diameters remain close to that of the initial alumina templates. Therefore, no correlation could be obtained between $r_s$, relative density, grain diameter and aspect ratio and more experiments would be needed. Nevertheless, there is an optimum $r_s$ at 5000-6000 °C/min, for our sintering conditions, where the ceramics obtained have high relative density, large grain diameter, and high aspect ratio.

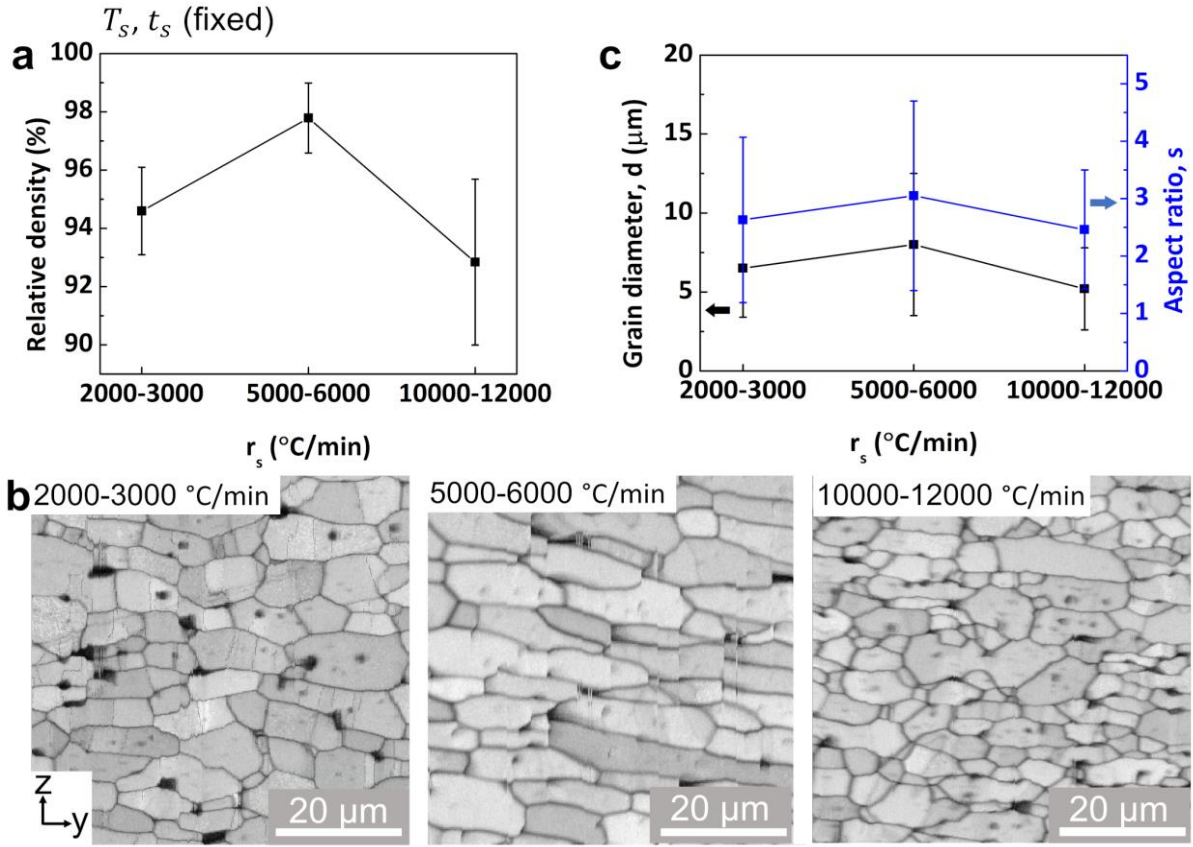

**Figure 5: Effect of heating rate $r_s$ on the density, microstructure and grain dimensions at a $T_s$ of 1700 °C and a $t_s$ of 10 s. (a)** Relative density as a function of $r_s$. **(b)** Band contrast images of the polished cross-sections showing grain surfaces (brighter regions) and grain boundaries (dark regions) in the z-y plane (see inset in **Fig. 1b** for plane definition). **(c)** Grain diameter (in black) and aspect ratio (in blue) as a function of $r_s$.

Although our experimentation with the three experiments (experiment 1 to experiment 3) could not explore all the temperature regimes, heating rates and sintering times, we could establish conditions at which high relative density with horizontally aligned anisotropic grain morphology can be obtained. We found that the ceramics could be densified with a sintering time as low as 10 s *via* TGG. Comparing the relative density and grain morphologies obtained, we observed that $r_s$ = 5000-6000 °C/min led to the highest relative density and grain aspect ratio at 1700 °C. Next, we validate our observations by varying again the $r_s$ and $T_s$ in the densification region, from 1600 to 1700 °C (**Fig. 6**). We found that for all sintering conditions, $r_s$ of 5000-6000 °C/min is the best heating rate for our alumina ceramic composition for achieving high relative density and grain aspect ratio.

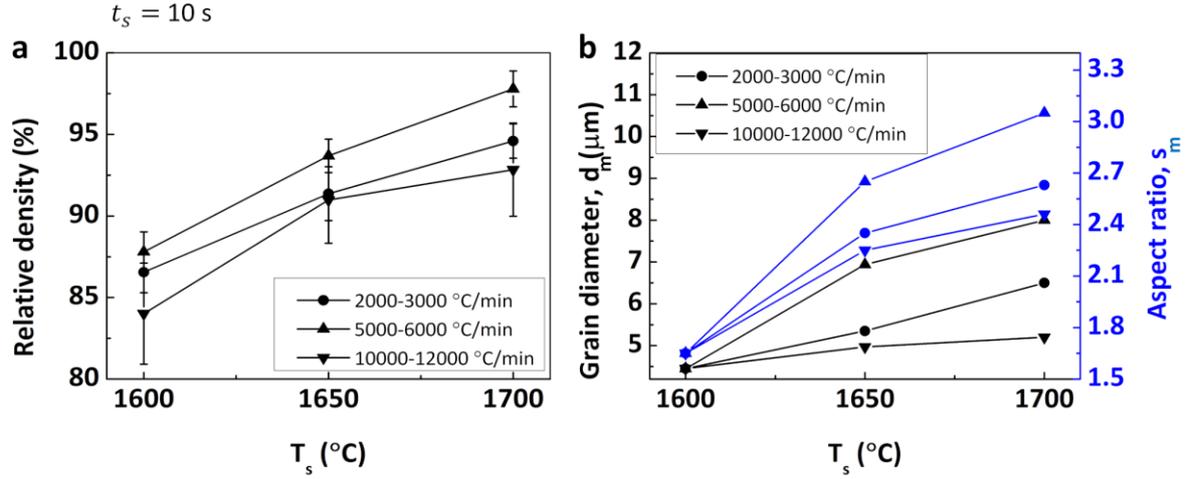

**Figure 6: Density and grain dimensions as a function of heating rate $r_s$ and sintering temperature $T_s$, at $t_s$ of 10 s. (a)** Relative density for the three different $r_s$ as a function of $T_s$. **(b)** Mean grain diameter $d_m$ (black) and mean aspect ratio $s_m$ (blue) as a function of $r_s$ and $T_s$.

The sintering conditions of $T_s$ = 1700 °C, $t_s$ =10 s and $r_s$ = 5000-6000 °C/min yielded the highest relative density ~98% and aspect ratio ~3 for grains of average diameter ~ 8 μm.

### 3.2 Sintering mechanisms to obtain TGG

In our material system, densification dominates grain growth at lower sintering temperatures (**Fig. 7a-c**). However, rapid TGG occurs for the three heating rates tested, leading to ceramics with a relative density of around 90% (**Fig. 7a-c**). This is similar to what was observed by Suvaci and Messing using uniaxial pressing followed by conventional sintering [9], in a composition with a similar initial size ratio between the template thickness and matrix grains of about 1.5, and a similar microplatelet concentration of 5-10 vol%. Furthermore, since densification precedes the grain growth of microplatelets, we can identify the point whereby there is densification with limited grain growth, also known as the frozen point [48]. For the three $r_s$, the frozen point of densification varies between 80 to 98%, with grain diameters of about 4.93-10 μm, depending on the heating rate (**Fig. 7a-c**). Furthermore, our results confirm that densification with meagre grain growth occurs predominantly at $T_s$ < 1700 °C, followed by rapid grain growth at $T_s$ ≥ 1700 °C. This mechanism is similar to what is observed in spark plasma sintering [34,49] and other sintering methods with high heating rates [35,50,51]. Overall, at $T_s$= 1700 °C, we obtain adequate grain growth with good anisotropy and densification, but for an ultra-fast sintering time at $t_s$= 10 s.

To have deeper insights into the TGG process occurring in UHS, we also compared the ceramics obtained after CS at 1600 °C to those obtained using UHS at different $T_s$ of 1600, 1650 and 1700 °C. TGG in CS was not observable at a $t_s$ of 2 h, but was found after sintering for around 10 h. In contrast, TGG in UHS was observed for a $t_s$ of 15 s only at 1600 °C, and for $t_s$ of 10 s at 1650 °C and 1700 °C (see **supplementary Fig. 8**). The ceramics obtained in CS and UHS in the conditions where TGG occurred had a relative density near to 90%. The delayed TGG in CS could be due to the less effective convective heat transfer in addition to the small amount of $Fe_3O_4$ used (0.75 vol% or 0.232 wt%) in an oxygen-rich atmosphere, which could have pinned the grains preventing their growth [43,44]. In contrast, the ultrafast TGG behaviour in UHS could be due to the efficient conductive heat transfer combined with the reducing sintering environment facilitating conversion of $Fe^{3+}$ to $Fe^{2+}$, leading to an ultrafast TGG within 10 s at 1650 and 1700 °C [42,52,53]. However, comparing the sintering mechanisms between the two processes is out of scope for this work.

In addition to ultra-short times for TGG, UHS also prevents the complete loss of low-temperature phases [19,28]. In our experimental protocol, we had coated the surface of alumina microplatelets with $Fe_3O_4$ nanoparticles that have a melting temperature of around 1600 °C. Elemental analysis of the samples after sintering at $T_s$ of 1700 °C for $t_s$ of 10 s shows concentration spots of Fe atoms (**Fig. 7d**). This is due to the ultrafast process whereby there is insufficient time for the Fe to diffuse completely. This is different from the CS process, where no concentration of Fe elements could be observed (see **supplementary Fig. 9**). The use of liquid phase precursors such as dopants and coatings are usually employed to reduce the sintering temperature and improve the densification in TGG; however, their presence after CS has not been confirmed [11,15,16]. In contrast, the presence of these precursors after sintering could be observed in the SPS [54,55], UHS [19,28], and other fast sintering processes [38], which could potentially improve the properties [56].

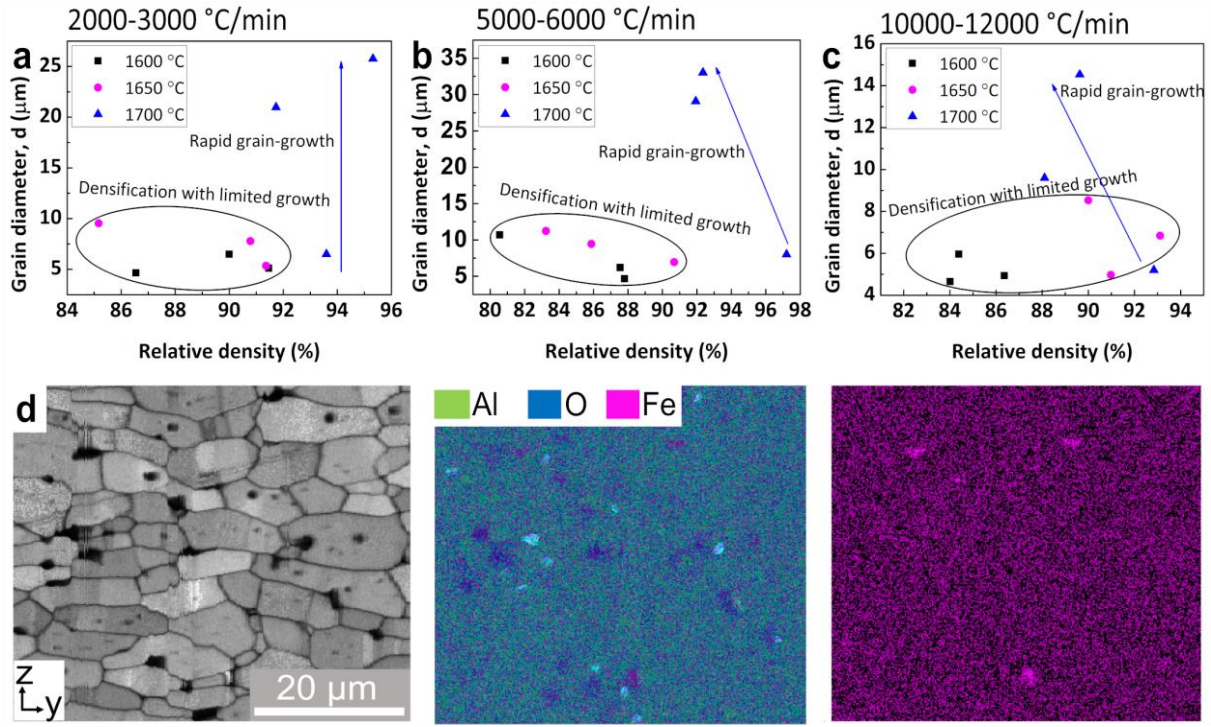

**Figure 7: Sintering mechanisms. (a-c)** Grain diameter as a function of the relative density for varying sintering temperature $T_s$ at heating rates of 2000-3000, 5000-6000 and 10000-12000 °C/min, respectively. These plots have two regions, the 'densification with limited growth' for $T_s$ <1700 °C (circled), and the 'rapid grain growth' at $T_s$ =1700 °C (shown by the arrow). **(d)** Elemental analysis of UHS sample obtained at 1700 °C for 10 s at 2000-3000 °C/min along the z-y plane. Left: Band contrast image showing the microstructure, where the brighter regions show grain surfaces (good pattern quality) and dark regions show grain boundaries (poor pattern quality); middle: Overall elemental distribution with the respective colour codes; right: Fe concentration spots.

## 3.3 Sintering map for obtaining textured alumina using UHS

To summarise the type of grain growth obtained for the sintering parameters we varied in UHS for our magnetically-aligned alumina grain bodies, we plotted the aspect ratio *s* as a function of the sintering temperature $T_s$ for the three heating rates tested $r_s$ for a sintering time $t_s$ of 10 s (**Fig. 8a**). In these plots, we can map three different temperature regions based on the type of grain growth observed. To have a more precise estimation for the prediction of the regions, we collected more electron micrographs at different temperatures and hence plotted the regions as shown in **Fig. 8a**. For $T_s$ < 1640 °C, we did not observe grain growth of anisotropic grains, and the matrix nanoparticles remained spherical nanograins. This region is labelled 'No TGG'. For 1640 °C ≤ $T_s$ ≤ 1780 °C, we observed TGG, yielding a dense ceramic with horizontally aligned grains, where the nanograin matrix is not visible anymore. For $T_s > 1780$ °C, the grains had a random and unusually large size with a non-uniform microstructure and the region is labelled 'AGG'.

Furthermore, to verify that our morphological texture with horizontally aligned anisotropic grains corresponds to the desired crystallographic texture, we used EBSD to obtain the crystallographic orientation map of the samples in the x-y plane (**Fig. 8b-d**, see **supplementary Fig. 10** for full EBSD scan). This x-y plane is the plane in which we aligned the basal or c-plane of the initial template particles, *i.e.* the magnetically aligned alumina microplatelets, in the green body. Therefore, after TGG, the orientation of the grains is along {0001} in the c-plane or x-y plane of the ceramic as observed. Also, at $r_s$ = 5000-6000 °C/min, the orientation along the x-y plane i.e., {0001} plane was observed to be the highest, as shown by the pole figure (**Fig. 8c**), in comparison to other $r_s$ (**Fig. 8b-d**). This is expected from our morphological study, and the large TGG observed at $r_s$=5000-6000 °C/min.

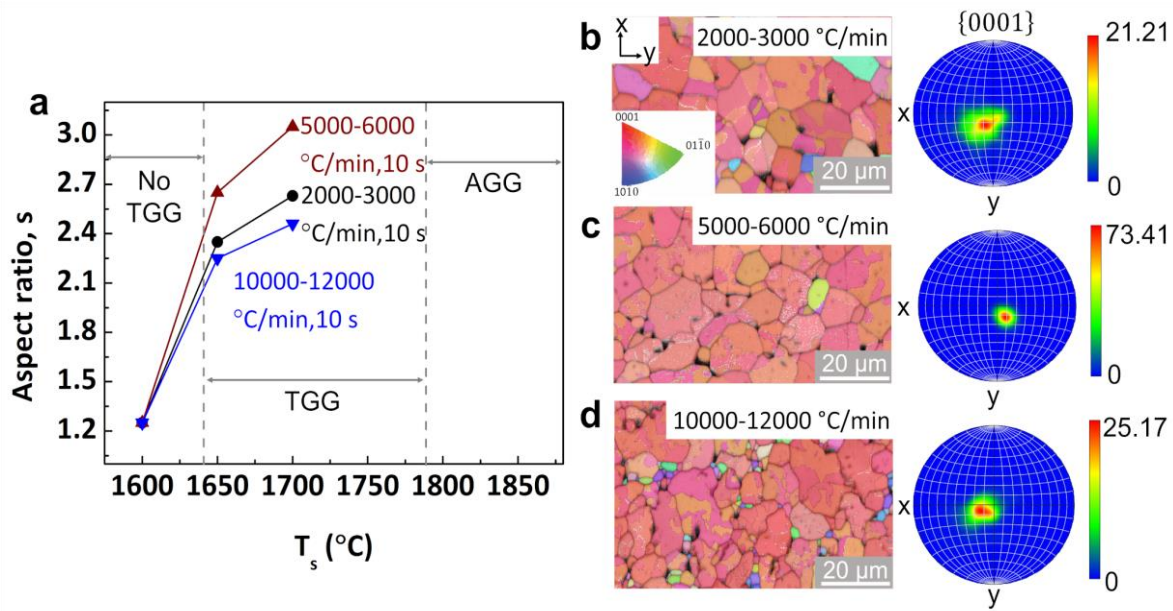

**Figure 8: Sintering map for obtaining dense alumina with crystallographic texture using MASC and UHS. (a)** Aspect ratio as a function of sintering temperature for a sintering time $t_s$ of 10 s, showing three different temperature regions: No TGG: No templated grain growth; TGG: Templated grain growth; and AGG: Abnormal grain growth. **(b-d)** Inverse pole figure (IPF) maps and the corresponding pole figures of the ceramics sintered in UHS at 2000-3000, 5000-6000 and 10000-12000 °C/min respectively, on the x-y plane, for a sintering temperature $T_s$ of 1700 °C and time $t_s$ of 10 s. The data shown in b-d are smaller sections from EBSD scans of larger areas containing many more grains (**supplementary Fig. 10**).

Our results validate that UHS can be used on magnetically-aligned alumina green bodies to yield dense and textured alumina ceramics within seconds. In the following, we measured the Young's moduli and hardness for the ceramics obtained to test for anisotropic behaviour.

## 3.4 Mechanical properties

The mechanical properties were tested on the UHS samples prepared at a sintering temperature of 1700 °C, for 10 s, for various heating rates $r_s$ (**Fig. 9)**. All samples exhibited anisotropic mechanical properties except those sintered at a $r_s$ of 10000-12000 °C/min. The moduli measured by nanoindentation on the z-y plane values were higher than on the x-y plane (**Fig. 9a**). This is in agreement with the fact that the plane perpendicular to the basal plane or c-plane is stiffer and harder than the plane parallel to it [1,12,13,57–60]. However, for 10000-12000 °C/min, the porosity, presence of many nanograins, inhomogeneities and misorientation, which were observed in electron diffraction images (see **Fig. 8d**, **supplementary Fig. 10**), tend to affect the modulus values, thus leading to loss of the expected anisotropic behaviour. Nevertheless, this loss in anisotropic behaviour is not observed in the micron-scale, and the Vickers hardness values agree well with the obtained texture, with no substantial difference in values ~12-13 GPa on an average between the different $r_s$ (**Fig. 9b**). Noticeably, the average Vickers hardness values of ~12-13 GPa are comparable to earlier reported values of similar dense aligned textured alumina obtained using CS (~12 GPa) [30], and SPS (13.97 GPa) [59]. However, the moduli obtained, varying from 210-275 GPa for our ceramics sintered using UHS, tend to be lower than the moduli of 400 GPa reported previously using SPS [59].

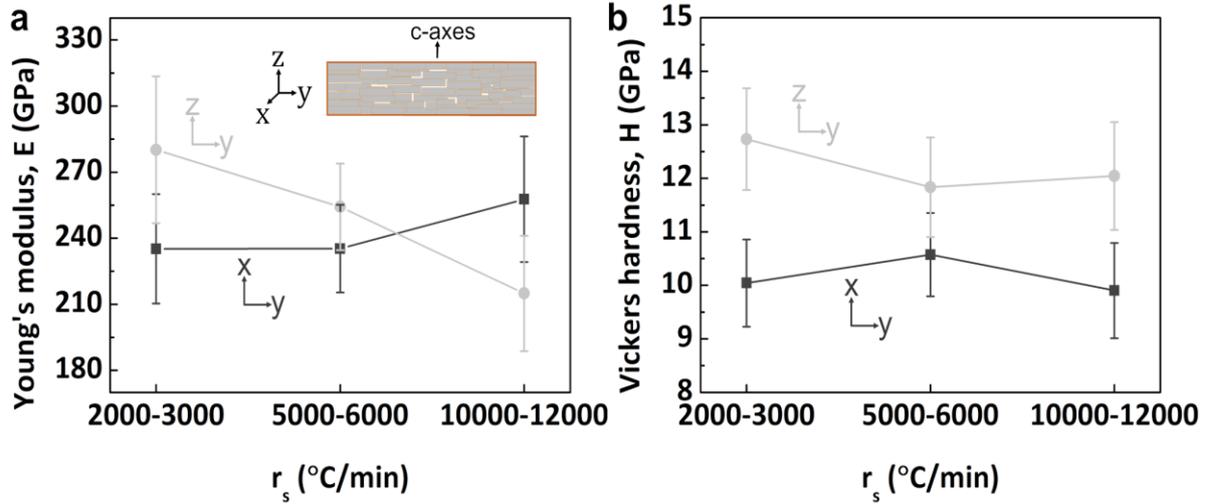

**Figure 9: Mechanical properties of the ceramics sintered at $T_s$ =1700 °C for $t_s$=10 s for different heating rate $r_s$. (a)** Young's modulus measured using nanoindentation as a function of the heating rate on the plane parallel to the basal plane or c-plane of the microplatelets (x-y), and perpendicularly (z-y). **(b)** Vickers hardness as a function of the heating rate on the planes (x-y) and (z-y).

## 4. Conclusion

In this work, we explored the use of UHS to sinter green bodies containing horizontally-oriented alumina microplatelets coated with $Fe_3O_4$ nanoparticles and dispersed in a matrix of alumina nanoparticles. We tuned the sintering parameters to find the conditions to obtain high relative density, anisotropic grains and crystallographic texture *via* TGG. Carrying out a systematic study of the sintering parameters, we found a narrow range where TGG occurs, i.e, $1640 \leq$ sintering temperature $T_s \leq 1780$ °C. The best sintering conditions were a sintering temperature $T_s$ of 1700 °C, a sintering time $t_s$ of 10 s for all the heating rates $r_s$. In these conditions, the Fe atoms from the $Fe_3O_4$ nanoparticles' coating did not diffuse completely, which contributed to maintaining the grain anisotropy. Furthermore, we observed densification to dominate at lower $T_s < 1700$ °C, while grain growth dominates at $T_s \geq 1700$ °C for all the heating rates $r_s$. Electron diffraction patterns with pole figures confirmed the crystallographic orientations for all $r_s$. An intermediate $r_s$ of 5000-6000 °C/min was found to yield the best crystallographic texture. Additionally, mechanical characterisation of the Young's modulus at nanoscale and the Vickers hardness at micron-scale indicated texture-dependent properties. Higher values were obtained perpendicular to basal plane than the parallel plane, except for 10000-12000 °C/min where the anisotropy was lost at the nanoscale. In conclusion, magnetic orientation followed by TGG in UHS is a simple, pressureless method to obtain highly textured alumina ceramics with anisotropic mechanical properties. This simple strategy could be applied to other ceramic systems to explore other functional crystallographic texture-dependent properties for a broad range of applications.

## Declaration of competing interests

The authors declare that they have no competing financial interests.

## Acknowledgements

The authors acknowledge Evonik Pte Ltd for providing samples of AERODISP® W 440. The authors acknowledge Facility for Analysis, Characterisation, Testing and Simulation (FACTS), Nanyang Technological University, Singapore, for the use of their electron microscopy and XRD facility. This research was funded by the National Research Foundation of Singapore (award NRF-NRFF12-2020-0002). The authors would like to acknowledge with thanks the financial support of the work by the project with PA number of DSOCL 21115, Singapore. We thank Ankit Jaiswal for helping us with the XRD.

## Data availability

The data that support the findings of this study are available from the corresponding author on request.

# Ultrafast high-temperature sintering of dense and textured alumina

Rohit Pratyush Behera[1], Matthew Jun-Hui Reavley[2,3], Zehui Du[3], Gan Chee Lip[2,3], Hortense Le Ferrand[1, 2*]

[1] *School of Mechanical and Aerospace Engineering, Nanyang Technological University, 50 Nanyang avenue, Singapore 639798*
[2] *School of Materials Science and Engineering, Nanyang Technological University, 50 Nanyang avenue, Singapore 639798*
[3] *Temasek Laboratories, Nanyang Technological University, 50 Nanyang Drive, Singapore 637553*

* Corresponding authors: hortense@ntu.edu.sg

Contents

**Supplementary note 1:**

The microcontroller is programmed to make the power supply output based on a user-defined current profile. Ramping and dwelling are defined in the current profile (**supplementary Fig. 1a**). Once the profile reaches the start of the dwell stage, the microcontroller takes a power measurement, which it then sets as the power ceiling (**supplementary Fig. 1b**). Subsequently, it continues to measure the power during the dwell stage, and reduces the current every time the power ceiling is exceeded in order to keep the output power just under the power ceiling. As observed in **supplementary Fig. 1a,b**, once the dwell stage is reached after 5s, the current occasionally reduces, and the output power reduces at that instant until it reaches the initial power ceiling and the current is reduced slightly again. As such, the power profile adheres to the programmed ramp up, dwell stage and ramp down.

The power profile data represents the output power from the power supply into the heating elements, and the temperature data comes from the measurements taken with the pyrometer (**supplementary Fig. 1c**). The primary discrepancy observed with this setup is a lag in temperature versus the power, but due to the high thermal conductivity of both the carbon heating elements and the sample, as well as the relatively low mass of both parts, the effect of this is minimal. As can be observed in **supplementary Fig. 1b,c**, the profiles are similar, with variation in temperature between the beginning of the dwell and the end being within 50 °C. Further, since the power profiles behave consistently, and the size of the samples and the heating elements are kept similar, this effect is consistent across samples.

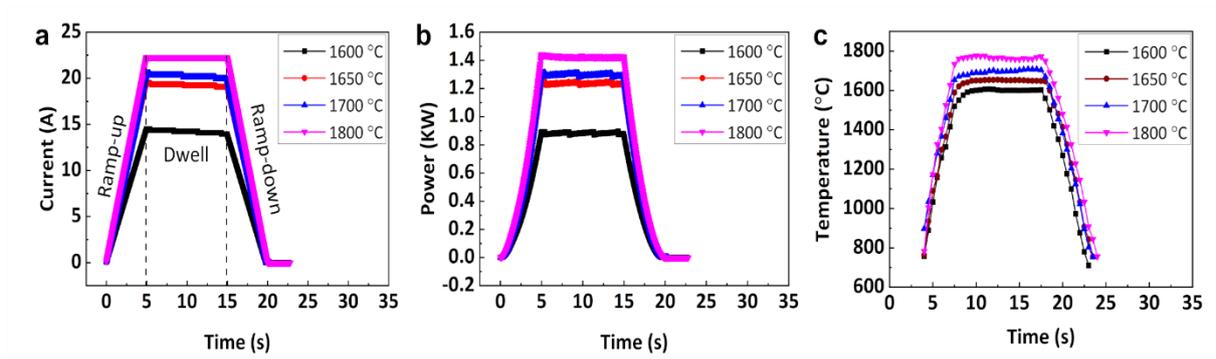

**Supplementary Figure 1**: (a) Supplied current (A) variation with time (s); (b) Regulated power (KW) controlled with the span of time (s); (c) Final obtained temperature (°C)-time (s) profile for average peak temperatures in the range 1600-1800 °C for a dwell time of 10 s.

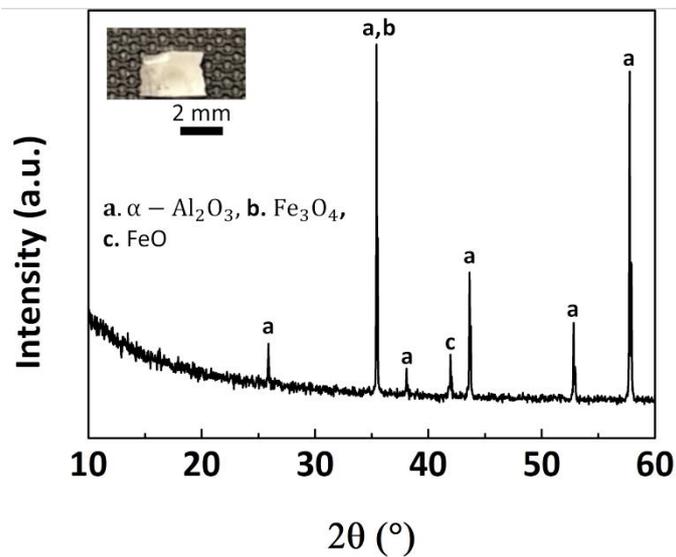

**Supplementary Figure 2**: **XRD of UHS sintered sample at 1800 °C**. The peak '**c**' is from the $Fe_3O_4$ added initially to magnetise our microplatelets.

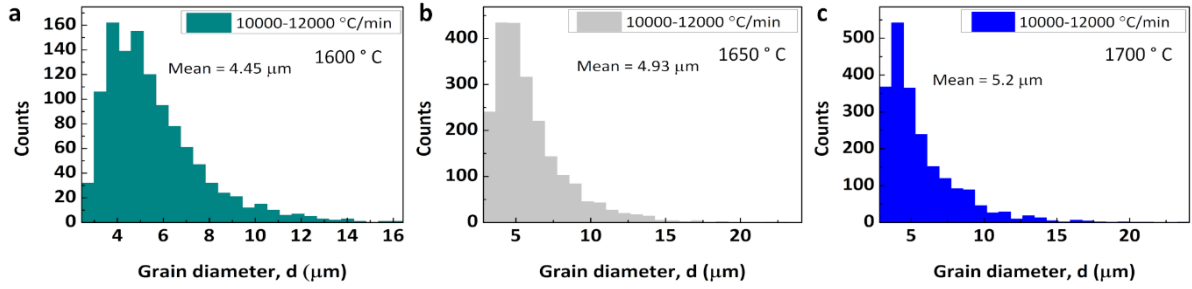

**Supplementary Figure 3**: Mean grain diameter estimation at $r_s$=10000-12000 °C/min at $t_s$= 10 s for different $T_s$ of (a) 1600 °C, (b) 1650 °C and (c) 1700 °C as obtained from EBSD.

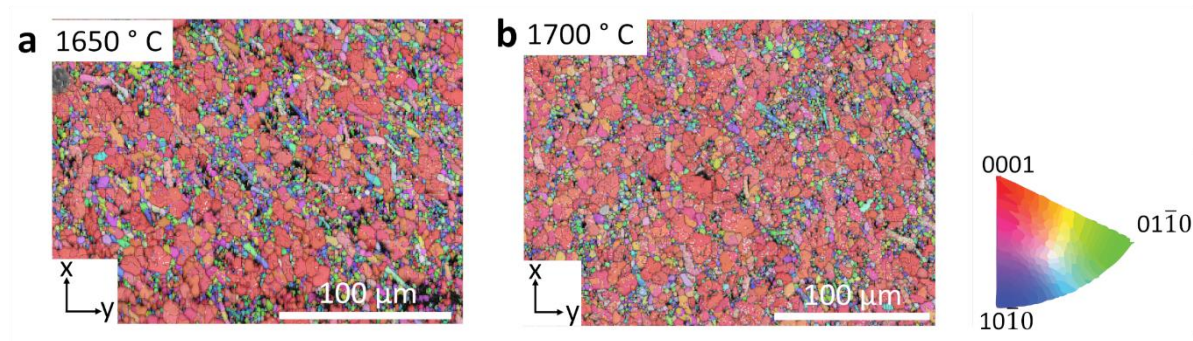

**Supplementary Figure 4**: Inverse pole figure (IPF) mapping from EBSD revealing grain boundaries and grain orientations for samples sintered at $r_s$=10000-12000 °C/min, $t_s$= 10 s and at a $T_s$ of (a) 1650 °C and (b) 1700 °C.

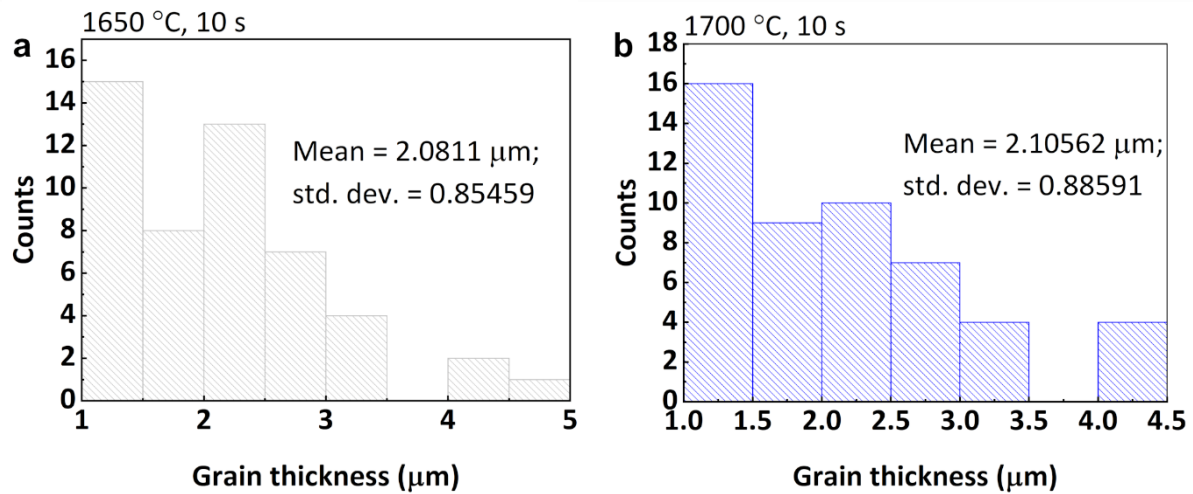

**Supplementary Figure 5**: Normal/Gaussian distribution plots for grain thickness of samples at $r_s$ =10000-12000 °C/min at 10 s for (a) 1650 °C and (b) 1700 °C.

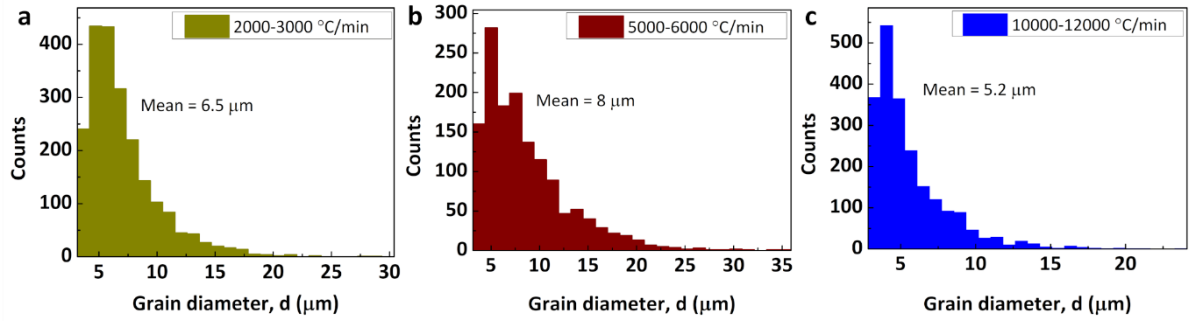

**Supplementary Figure 6**: **Grain diameter of samples for $T_s$= 1700 °C, $t_s$= 10 s at different $r_s$ of** (a) 2000-3000 °C/min, (b) 5000-6000 °C/min, and (c) 10000-12000 °C/min as obtained from EBSD.

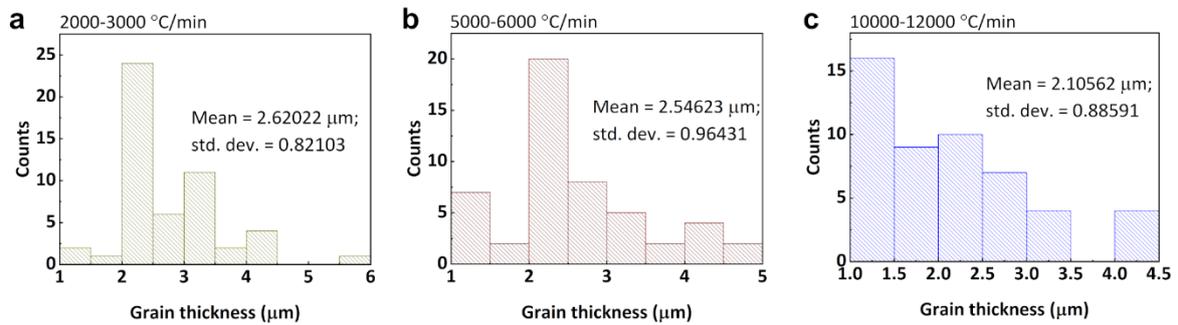

**Supplementary Figure 7**: **Normal/Gaussian distribution plots for grain thickness of samples sintered at 1700 °C for 10 s at different $r_s$ of** (a) 2000-3000 °C/min, (b) 5000-6000 °C/min and (c) 10000-12000 °C/min.

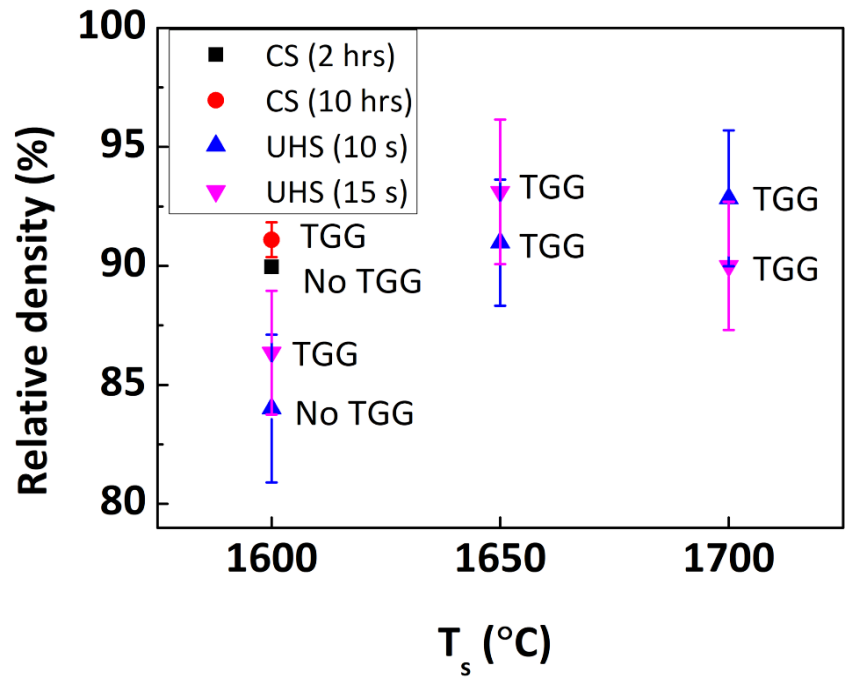

**Supplementary Figure 8**: TGG process comparison of the conventionally sintered (CS) and UHS samples.

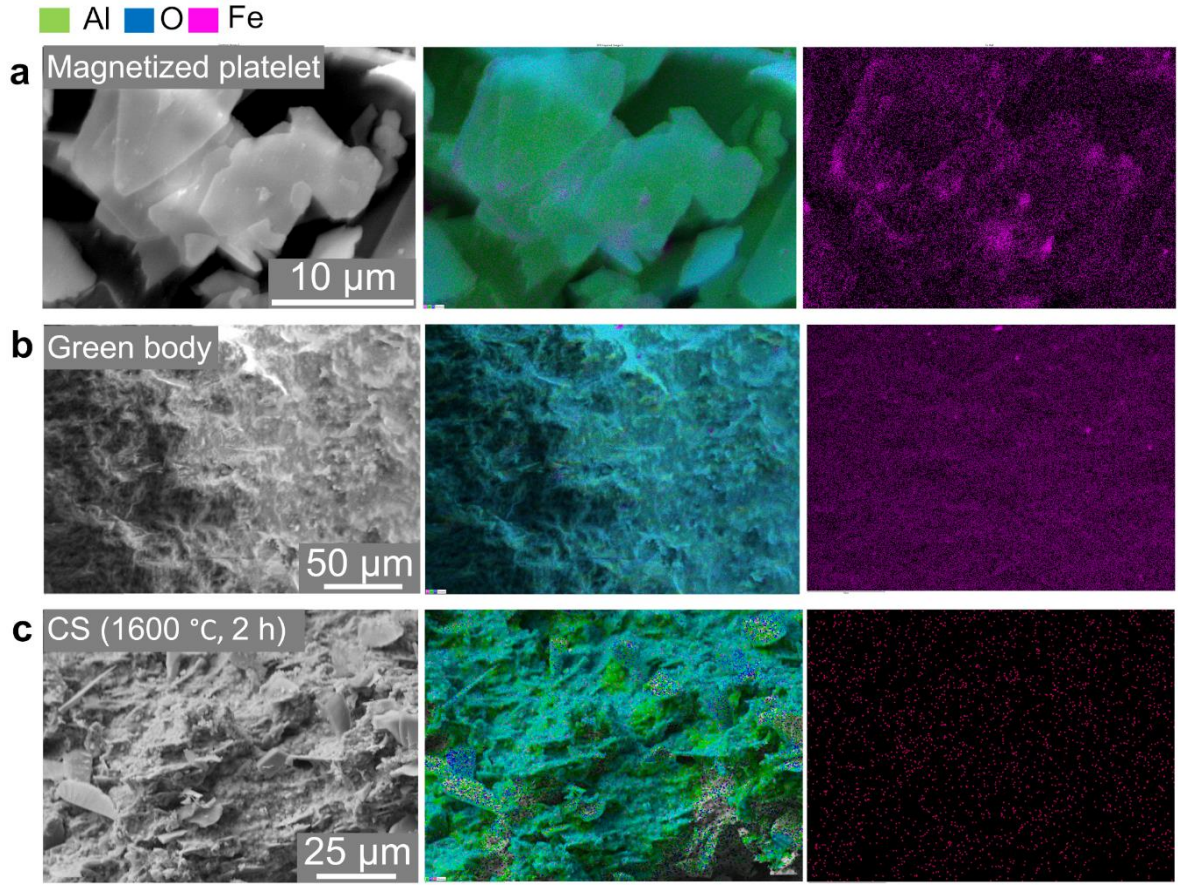

**Supplementary Figure 9**: **Elemental mapping showing the distribution of 'Fe'**. Electron micrographs followed by overall elemental distribution and 'Fe' distribution for the (a) magnetised platelet, (b) green body, and (c) the CS sample at 2.5 °C/min, 1600 °C for 2 h.

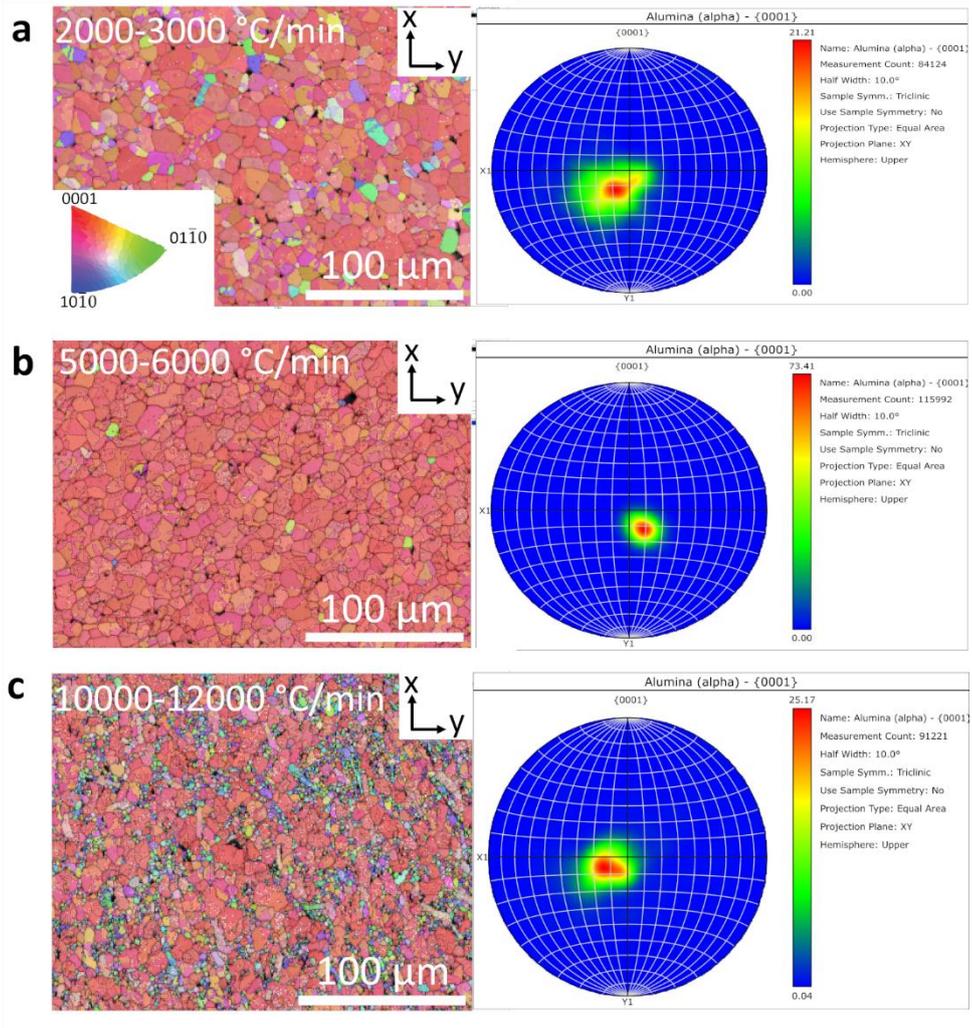

**Supplementary Figure 10**: **Crystallographic mapping of UHS samples showing large scan area at $T_s$ = 1700 °C and $t_s$ = 10 s for the different $r_s$ of** (a) 2000-3000 °C/min, (b) 5000-6000 °C/min and (c) 10000-12000 °C/min.